\documentclass{statsoc}
\usepackage{amssymb,fullpage,epsfig,color}
\usepackage[dvipsnames]{xcolor}
\usepackage[]{graphicx,mathtools}

\title[Stats didn’t prove Seafood Market was COVID origin]{Statistics did not prove that the Huanan Seafood Wholesale Market was the early epicenter of the COVID-19 pandemic\\[1ex]
\underline{\rm\small Running head:} {\it\small Stats didn’t prove Seafood Market was COVID origin\/}}
\author[Stoyan and Chiu]{Dietrich Stoyan}
\address{Institut f\"{u}r Stochastik, TU Bergakademie Freiberg, Freiberg, Germany}
\email{stoyan@math.tu-freiberg.de\\}

\author{Sung Nok Chiu\thanks{Corresponding author}}
\address{Department of Mathematics, Hong Kong Baptist University, Kowloon Tong, Hong Kong}
\email{snchiu@hkbu.edu.hk}

\begin{document}

\begin{abstract}
In a recent prominent study Worobey et al.\ (2022, Science, 377, pp.\ 951--9) purported to demonstrate
statistically  that the Huanan Seafood Wholesale Market was the epicenter of the early COVID-19
epidemic. We show that this statistical conclusion is invalid on two grounds:
(1) The assumption that a centroid of early case locations or another
simply constructed point is the origin of an epidemic is unproved.
(2) A Monte Carlo test used to conclude that no other location than the seafood market can be the origin is flawed. Hence, the question of the origin of the pandemic has not been answered by their statistical analysis.
\end{abstract}


\section{Introduction} \label{sec:1}
{\it On 31 December 2019, the Chinese government notified the World Health Organization (WHO) of an
outbreak of severe pneumonia of unknown etiology in Wuhan, Hubei province.} This may be considered
the beginning of the COVID-19 pandemic. Soon the question of
its origin was asked and since then discussed controversially. Currently, in 2023, there are two
main hypotheses: (a) there is a zoonotic origin, the virus came from animals (the zoonosis hypothesis);
and (b) there was an accident in a laboratory,
somehow the virus fled from human supervision (the `lab-leak' hypothesis).  Each hypothesis is, however, not necessarily the alternative of the other.

On 26 July 2022 {\it Science\/} published the paper \citet{Worobeyetal2022}, which says clearly in its title
``{\it The Huanan Seafood wholesale Market in Wuhan was the early epicenter of the COVID-19 pandemic}''
and in its abstract: ``{\it We show here that the earliest known COVID-19 cases from December 2019, including those without reported direct links, were geographically centered on this market\/}''.
This paper, to which we refer in the following as {\bf W}, has attracted world-wide attention and media coverage and has been downloaded almost four hundred thousand times in 10 months after publication.
The paper uses for the proof of its title's statement two different arguments: a statistical one and a zoonotic one based on a coincidence, using the fact that in the seafood market animals (mammalia) are sold.

The statistics in {\bf W} mainly use {\it center-points\/}, which are defined by the coordinate-wise median latitudes and longitudes (see Supplementary Materials of {\bf W}).  {\bf W}'s use of the center-point to identify the ``center'' of a point cloud is analogous to the use of the median to measure the central tendency of a set of numerical data, obviously under the (unestablished) assumption that the ``center''
of a cloud of locations of cases starting from an origin of the infection process is close
to this origin.   However, using the coordinate-wise median to define the ``center'' of a point cloud is a questionable choice, because the coordinate-wise median is not rotationally invariant \citep{Walker2022a}. Using the centroid, which is the coordinate-wise mean, may be geometrically more  reasonable.  Another rotationally invariant representation of the ``center'' of a point cloud is the peak, or the mode, of the underlying spatial density function, if it is unimodal.
Moreover, {\bf W} considers only one location, namely, the seafood market, as a possible origin based on some non-statistical argument.  They carried out a Monte Carlo test, found that the median distance between the seafood market and the confirmed cases is significantly shorter than the median distance between the seafood market and independent points following the Wuhan population density, and concluded that because the confirmed cases are not random points distributed according to the population density, the seafood market was the origin of the infection process.

Their unconvincing analysis has led us to a critical view of the paper.
We will discuss with technical details
only the above-mentioned statistical aspects of
{\bf W} and will not judge their zoonotic argument. {\bf W} also reported their spatial relative risk analysis of environmental samples taken inside the seafood market, but in this paper we criticise only their arguments in finding the epicenter of the outbreak in Wuhan City; the analysis of the spatial data within the seafood market is hence irrelevant here.

The present paper is organised as follows.  We will comment on the problems in {\bf W}'s analysis in Section~\ref{sec:2}, and then we will focus on their spatial statistics methods. In Section~\ref{sec:3} we construct confidence regions for the location of the ``center'' of the point cloud of early cases by bootstrapping, while in Section~\ref{sec:4} we discuss the flaws in the test of hypotheses in {\bf W} and show how the hypothesis that the Market is the ``center'' of the point cloud of early COVID-19 cases actually should be rejected by a proper Monte Carlo test. Moreover, even if the seafood market could be established as the ``center'', causal inference would still be unjustified (see e.g.\ \citeauthor{DablanderHinne}, \citeyear{DablanderHinne}); centrality does not imply causality \citep{Walker2022b}.  Nevertheless, we acknowledge that sometimes it could be helpful to identify the ``center''.   In Section~\ref{sec:5} we review the approaches adopted in two classical studies for identifying the source of an outbreak, discuss why these approaches may not be applicable to the COVID-19 data, and suggest some possible directions for statistical analysis if data of better quality were available.

The discussion in the following sections will lead to the conclusion that their statistics arguments are not convincing and do not provide sufficient evidence supporting the claim that the Market was the early epicenter.
Our disapproval of {\bf W}'s approach does,
of course, not mean that we reject the zoonosis hypothesis; we just consider the question
of which of the two hypotheses is true has not been answered by the analysis of these spatial data.

\section{General remarks on the paper {\bf W}} \label{sec:2}
The starting point of the statistics in {\bf W}, as well as in the present paper, is the point
pattern of address locations of the people infected in December 2019 (or ``cases'' for short). {\bf W} recovered 155 cases from the 164 cases shown in Annex~E2 Figure~4 of WHO Report \citep{WHO2021}.

Unfortunately, these data are of poor quality.  We identified at least four major problems.  First, the precise latitude and longitude coordinates of these locations are not available; {\bf W} claimed that the extraction method producing these data introduced no more than 50$\,$m of noise in each case. Second, there is a cluster of seven cases with the same address; these multiple locations are considered different in {\bf W}. Third, no onset date per case is known and used; even the outset of the beginning of the pandemic
was shrouded in uncertainty; Figure~1 in \citet{Holmes_etal2021}  gives
a vague impression of what could be possible if temporal data were also available. Fourth, the data are only partial; \citet{Demaneuf2022} reported that $257$ cases for December 2019 have been attested in papers such as
\citet{Shi2021}, resulting from a retrospective search that ended in February 2020.

However, this paper aims to critique not the source and quality of the data but the applied statistical
methods. We will still use {\bf W}'s data here, so that we will have the same point of departure.  Nevertheless, instead of using latitude and longitude coordinates and the Haversine distance, we project these locations to UTM coordinates, as shown in
Fig.~\ref{fig:1}, and use the Euclidean distance. The data and the R program used for the analysis can be found in supplementary materials of this paper.

\begin{figure}[htp]
\centering
\begin{tabular}{cc}
\hspace{-2.2em}\includegraphics[width=0.55\textwidth]{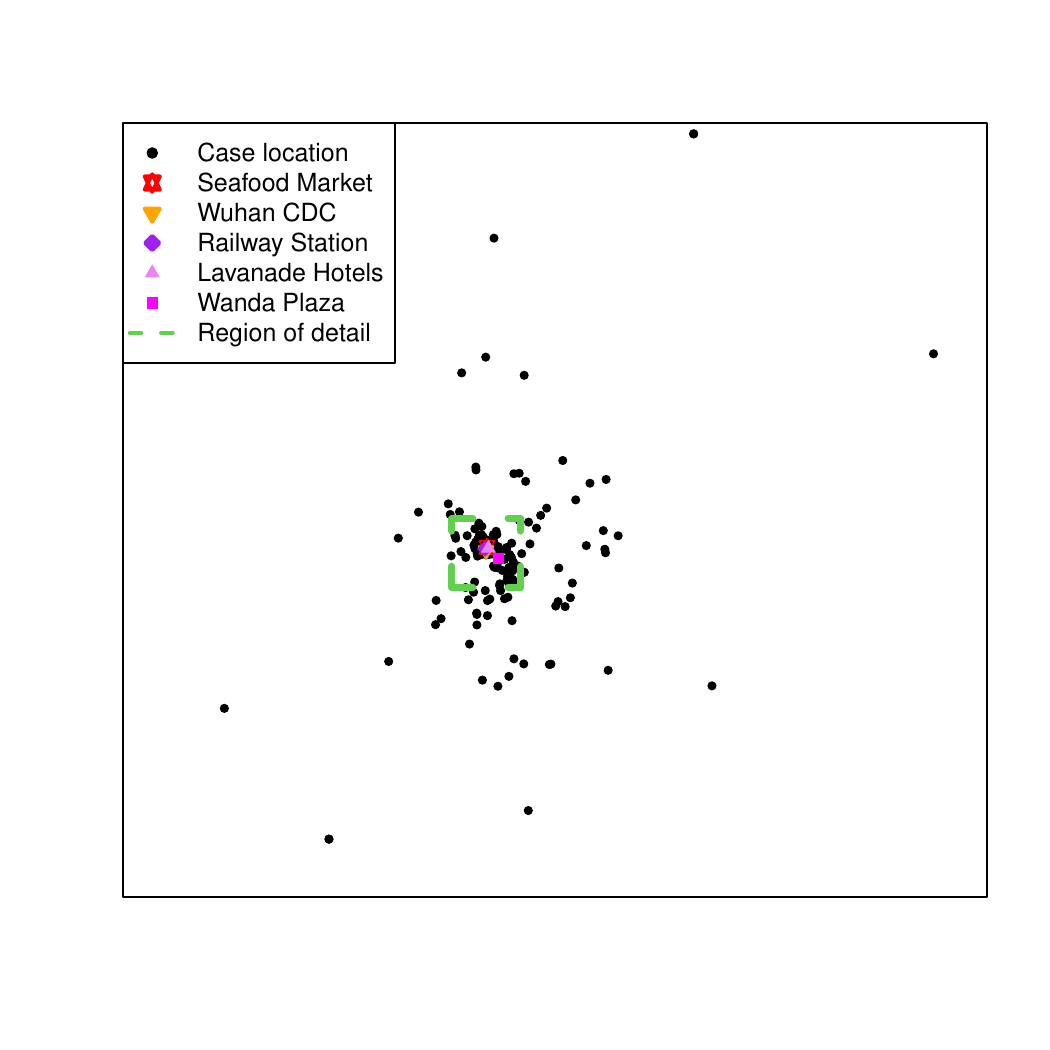}
&
\hspace{-2.2em}\includegraphics[width=0.55\textwidth]{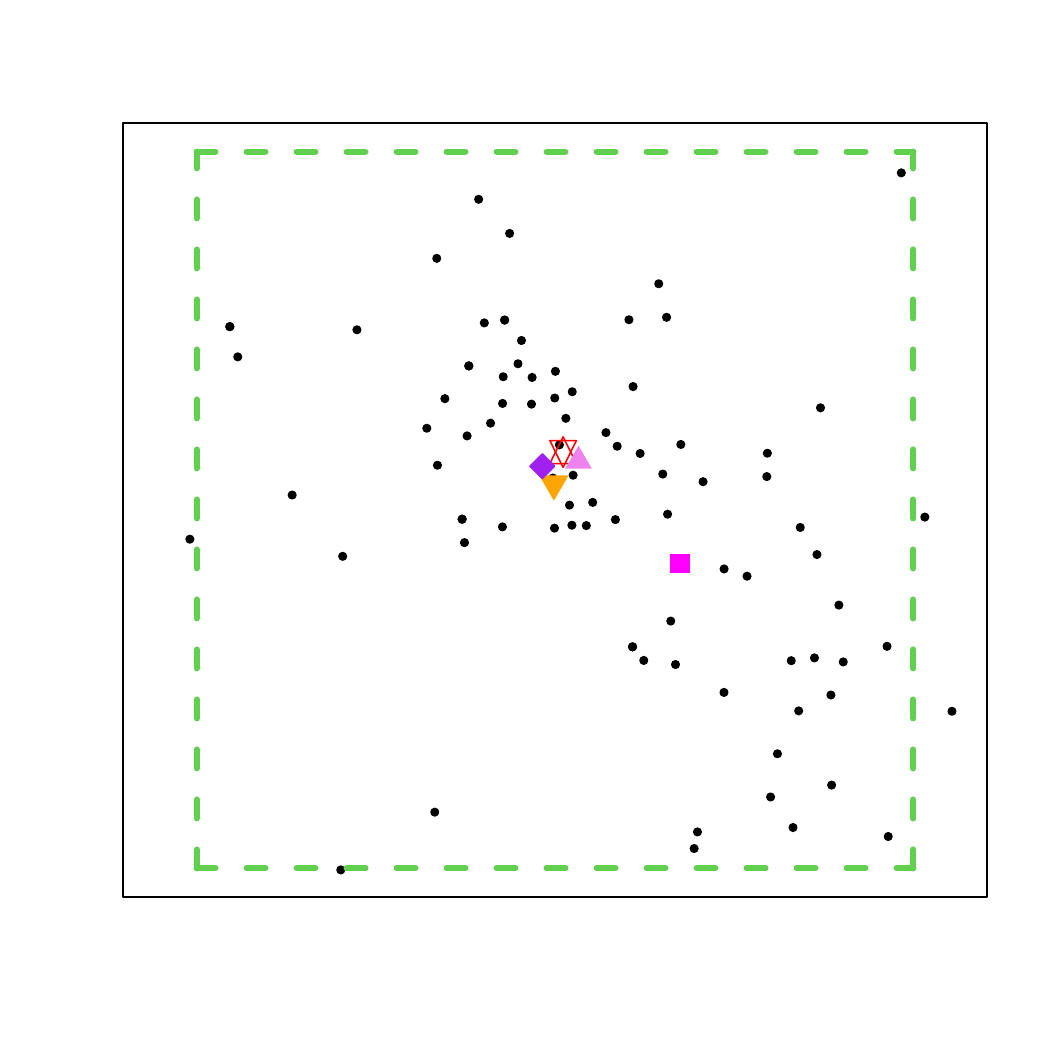}
\\[-2em]
\hspace{-1em}\footnotesize (a) & \hspace{-1em}\footnotesize (b)
\end{tabular}
\caption{\label{fig:1}(a) The 155 address locations of the people infected in December 2019, and some landmarks that are possible ``centers'' of the point cloud formed by the address locations. (b) The region of detail marked in (a).}
\end{figure}

\begin{figure}[htp]
\centering
\includegraphics[width=\textwidth]{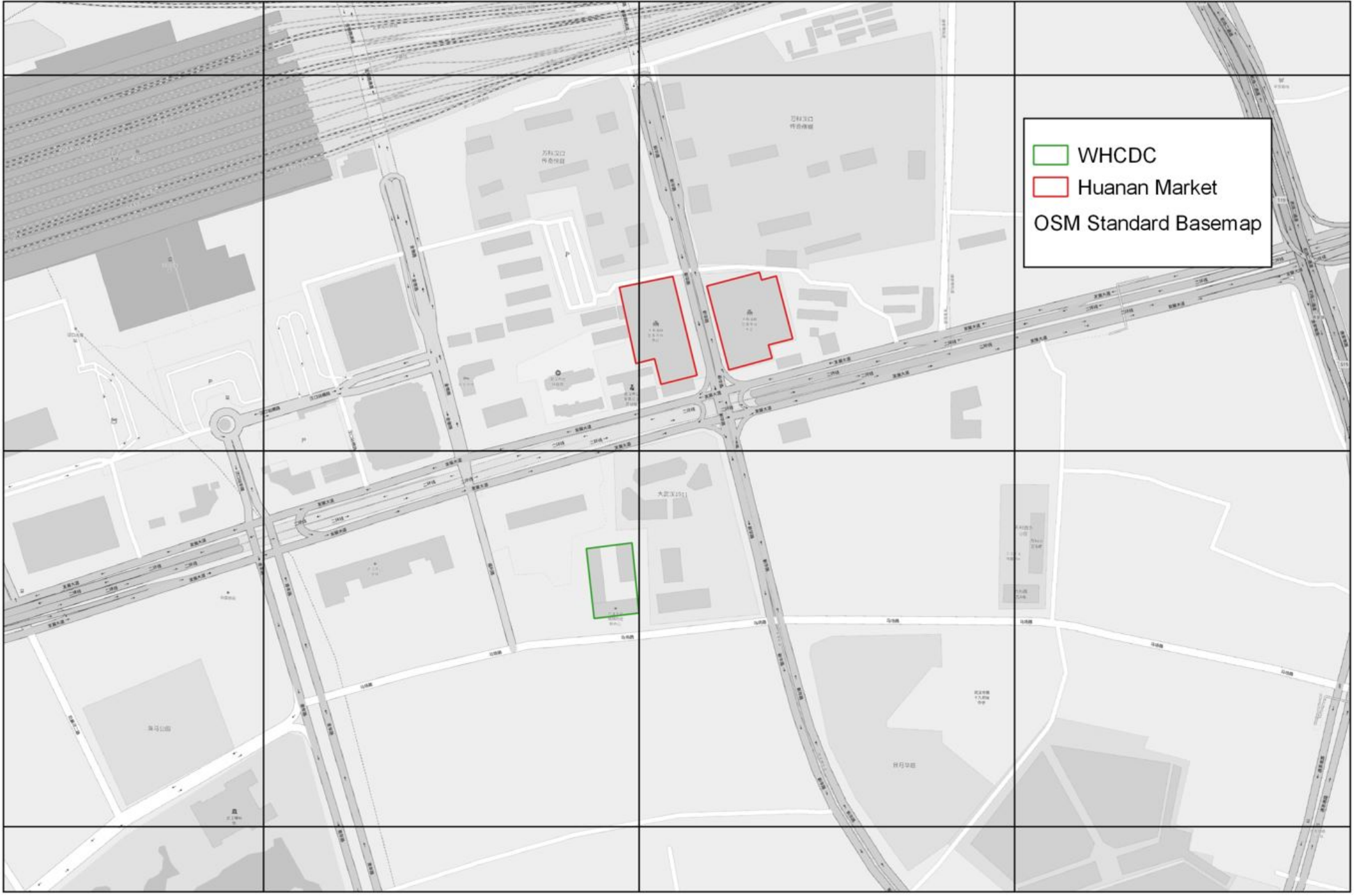}
    \caption{\label{fig:2}The Huanan Seafood Wholesale Market, the Wuhan Center for Disease Control and Prevention, and the Hankou Railway Station in UTM coordinate system.}
\end{figure}

{\bf W} considers the Huanan Seafood Wholesale Market (hereinafter the ``Market'') the ``epicenter'' of the pandemic, but near the Market there are various landmarks, which we will consider as
alternative possible candidates of the ``center'' of the point cloud of December cases and mark them in Fig.~\ref{fig:1}.  These include the Wuhan Center for Disease Control and Prevention (CDC), the Hankou Railway Station, and the Wanda Plaza, which is a shopping area with hotels and restaurants, near the Lingjiao Lake and its Park; in addition, one of the hotels listed in {\bf W} is also marked in Fig.~\ref{fig:1}.  Fig.~\ref{fig:2} is a map in UTM coordinate system that shows the relative sizes of and distances between the Market, the Wuhan CDC, and the Hankou Railway Station.  We follow {\bf W} and use points to represent the Market, as well as the landmarks.  Because of the presence of noise in {\bf W}'s location data of cases, the physical sizes of these landmarks could also be considered as noise in locations.

Note that we do not hypothesise that any of these landmarks is the origin of the pandemic, but only mark them in the plots as alternative possible ``centers'' of the point cloud of the 155 cases in our analysis to show that in context of statistics, the Market is not more likely to be the origin than the others are.  However, all of these landmarks are just hand-picked, and they do not form an exclusive list of all potential origins.  {\bf W} excluded all landmarks because they claimed that {\it no other location except the Huanan market clearly epidemiologically linked to early COVID-19 cases\/}.  In other words, according to {\bf W}'s approach, if epidemiological links can be found between the cases and any of these landmarks, then these landmarks will be equally likely to be the origin of the pandemic. Moreover, we should note that {\bf W}'s argument (on page~3 of 9) for the so-called direct linkage of the two lineage~A cases to the Market is that the first case was 2.31\,km away from the Market, significantly shorter than the distance from the Market to a random point generated according to the population density, while it was reported that the second case had stayed in a hotel near the Market; because there were at least 20 hotels within 500\,m of the Market, {\bf W} claimed that the second case could not have been more distant from the Market than the first case was.   The argument for direct linkage of the first case is obviously invalid because under this argument any locations in the vicinity of the Market, such as the Hankou Railway Station, were also epidemiologically linked.  The argument for direct linkage of the second case is seriously biased. Wuhan is a city of population 11 million and of size 8,500\,km$^2$, and has more than a thousand hotels; we could not see any scientific justification for {\bf W}'s claim that a hotel near the Market would mean a hotel not more than 2.31\,km away from the Market.  However, we do not refute the possibility that the Market is epidemiologically linked to the cases.

\section{Centroids, center-points and modes} \label{sec:3}

In the given data situation, it is a natural approach for the search
of a plausible origin of the pandemic to construct some ``center''
of the point cloud of cases. This may imply a nice, but perhaps overly
simplified, model of radial diffusion in an isotropic medium or a medium
with elliptical isolines.

{\bf W} follows this approach with the justification in the words of its authors
(line~6 on page~5 of Supplementary Materials of {\bf W}): {\it insofar as the center-point
of early cases might reflect the starting point of the epidemic.\/} As we mentioned in Section~\ref{sec:1},
{\bf W} uses as ``center'' the coordinate-wise median, referred to as  ``center-point'', but in this paper we consider also
the coordinate-wise mean, called the ``centroid'', and the peak, or ``mode'', of a kernel estimate of point density for
the point cloud of the cases.  Such a kernel estimate is given in {\bf W}'s Figure~1, which shows that the Market lies within the density contour enclosing a region of highest 1\% probability mass. However,
its uncertainty is not quantified and it is not mentioned that the Hankou Railway Station and the Wuhan CDC lie at positions with similar estimated probability
densities.

For all three choices of the representation of the ``center'' of the cases, namely, the center-point, the centroid and the mode, we do not have any statistical argument to propose that they are
valid estimates of the pandemic origin.  Instead, we believe that this is in fact not a question that these spatial data can answer.
However, we try to characterise the uncertainty of these constructed points representing the ``center'', in
order to evaluate the conclusions in {\bf W}.

For this purpose we consider confidence regions based on resampling of the pattern
of the 155 cases. Furthermore, we consider resampled patterns of smaller point numbers
with the aim to get, in another way, information of the variability of constructed
centers.

The duality between a confidence region and a hypothesis test should be noted here.  The Market would be inside the confidence region if and only if we accept the null hypothesis that the Market was the ``center'' of the early cases; however, this acceptance would only mean that there is insufficient evidence against this null hypothesis. Aiming at conclusions that could be termed ``statistically significant'', {\bf W} constructed a different null hypothesis and reported some small $p$-values in order to conclude that there is significant evidence that the Market was the early epicenter.  We consider their hypotheses inappropriate and the tests dubious; more detailed comments are given in the next section.

We take $m$ resamples, each of size $n \le 155$, sampled with replacement from the original 155 cases and determine the centroid, the center-point and the mode of a kernel density estimate of each resample.  In fact, we also considered sub-sampling without replacement, which gave us smaller confidence regions and smaller Monte Carlo $p$-values than bootstrapping; but we do not report the sub-sampling results here. The density function of cases was estimated by using circular Gaussian kernel with bandwidth 3\,km.

\begin{figure}[htp]
\centering
\begin{tabular}{cc}
\hspace{-2.2em}\includegraphics[width=0.55\textwidth]{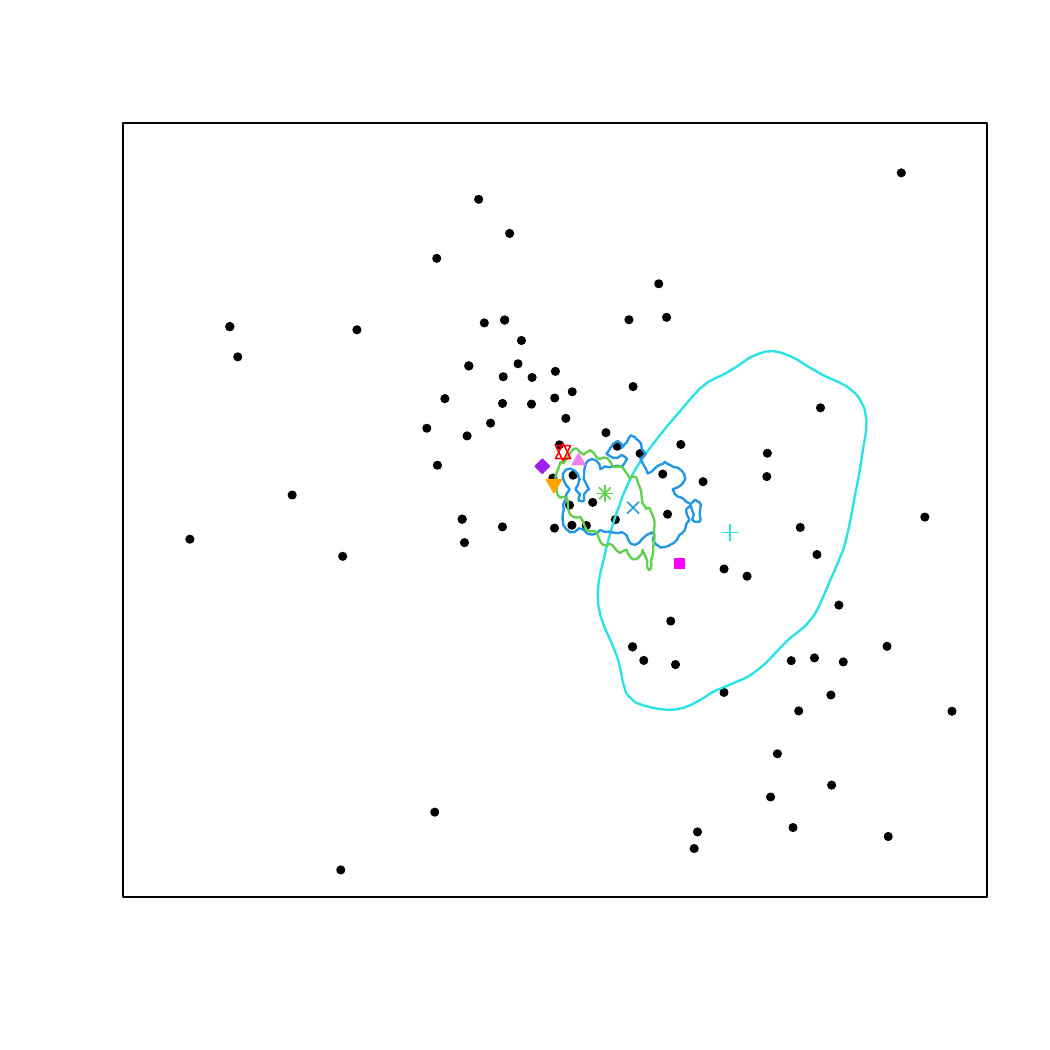}
&
\hspace{-2.2em}\includegraphics[width=0.55\textwidth]{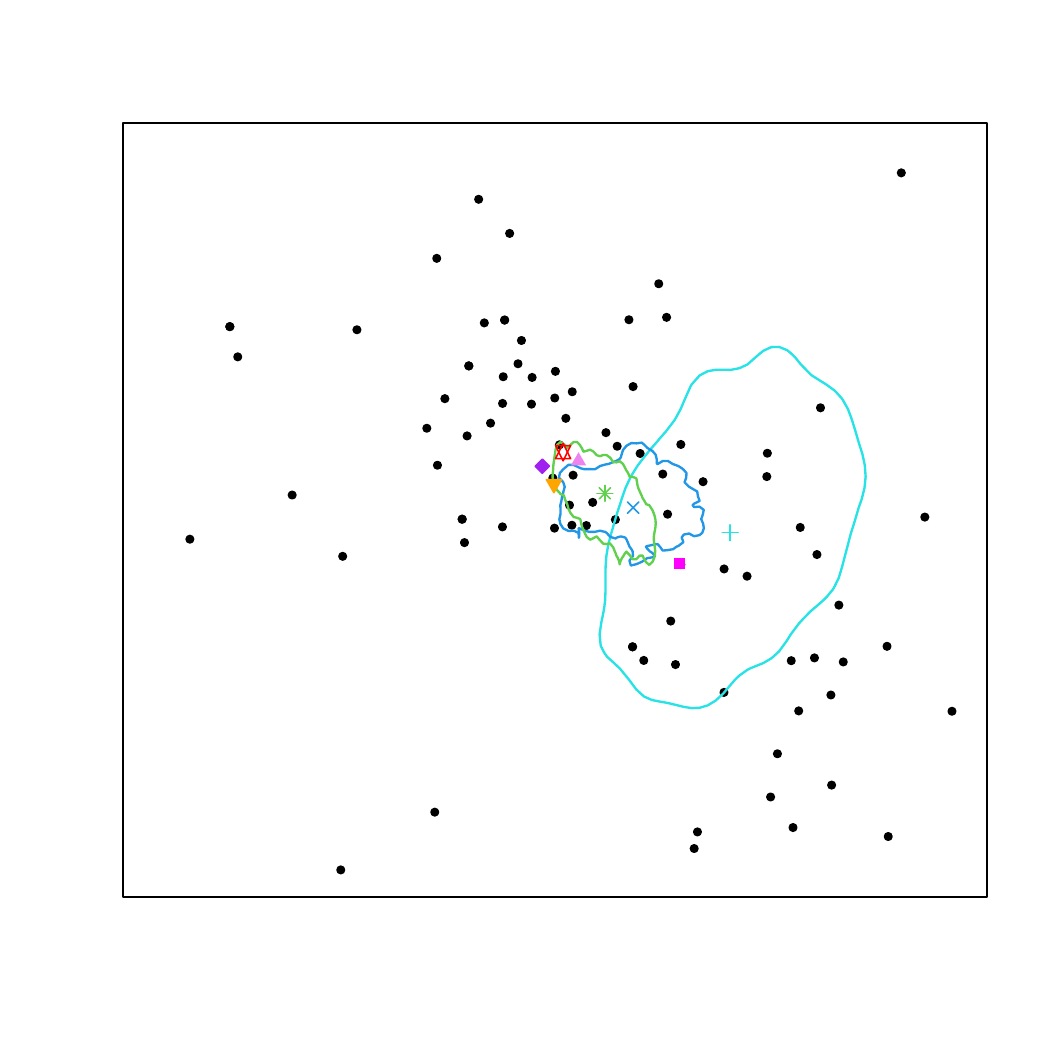}
\\[-2em]
\hspace{-0.75em}\footnotesize (a) $n=155$ & \footnotesize \hspace{-0.75em} (b) $n=150$\\
\hspace{-2.2em}\includegraphics[width=0.55\textwidth]{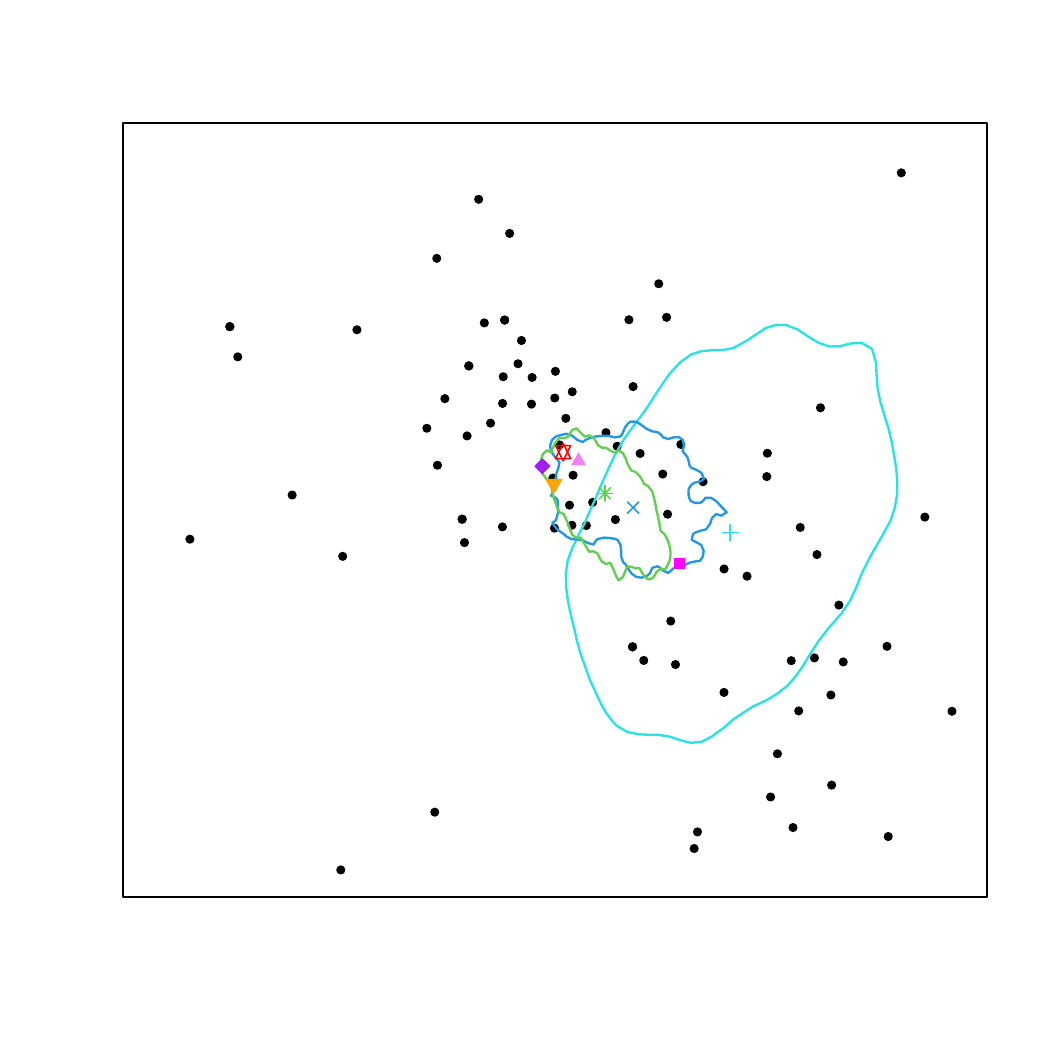}
&
\hspace{-2.2em}\includegraphics[width=0.55\textwidth]{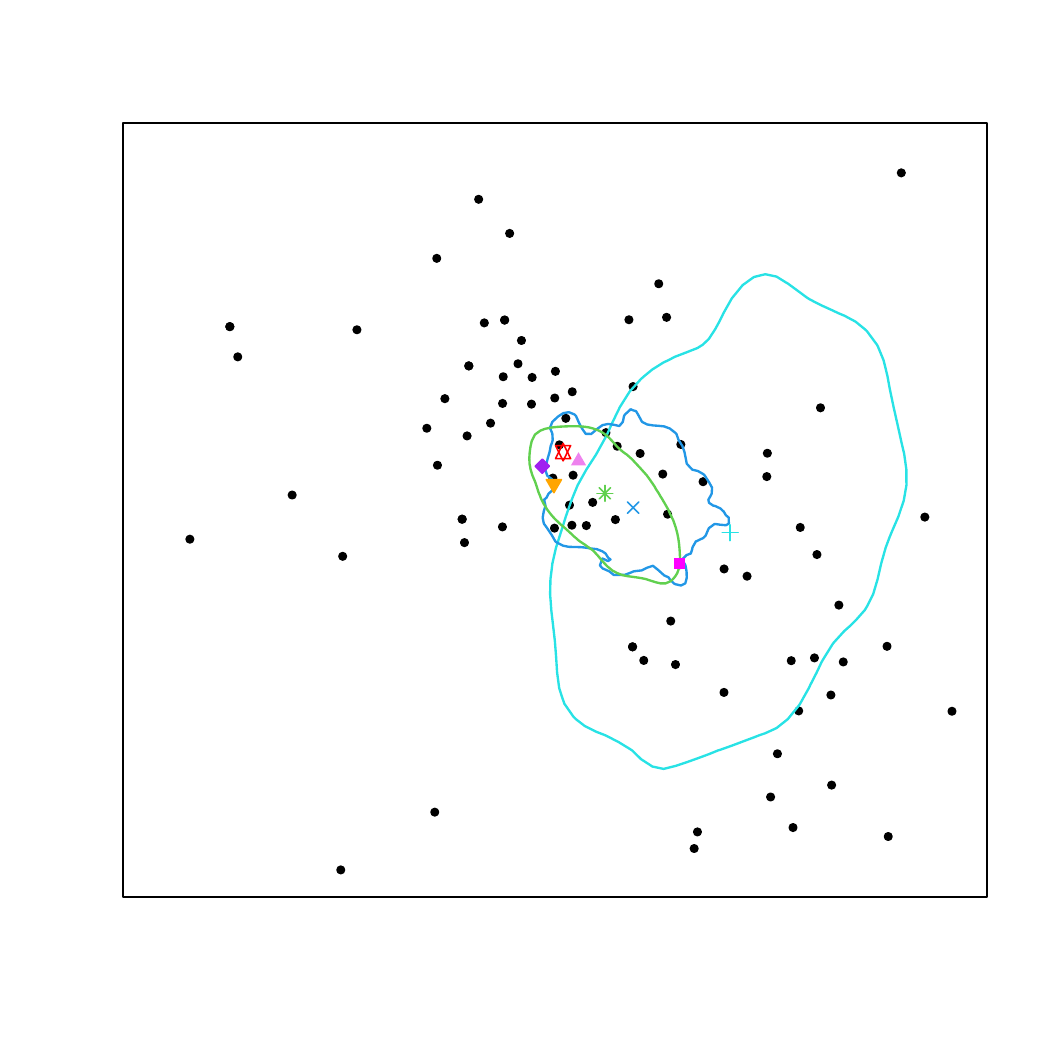}
\\[-2em]
\hspace{-0.75em}\footnotesize (c) $n=100$ & \footnotesize \hspace{-0.75em} (d) $n=80$\\
\end{tabular}
\caption{\label{fig:4}  The Market ({\color{red}${\mathrlap{\raisebox{1.2pt}{$\bigtriangledown$}}{\bigtriangleup}}$}), as well as
the centroid ({\color{cyan}+}), the center-point ({\color{blue}{$\times$}})
and the mode ({\color{green}{$\mathrlap{\raisebox{0pt}{$\times$}}+$}}) of all 155 cases ($\bullet$), and their 95\% confidence regions of 999 resamples with replacement of size $n$ in the region of detail. }
\end{figure}

The bootstrap results for $m=999$ and $n=155$, 150, 100 and 80 are shown in Fig.~\ref{fig:4}.  As expected, the smaller the size $n$ of the bootstrap samples, the larger the clouds of ``centers'' and hence the larger the resultant 95\% confidence regions.  A motivation for considering smaller sizes came from the fact that the original 155 cases are unlikely independent in reality, and any attempt to account for non-independence would likely give larger confidence regions. Therefore we considered smaller $n$ to enlarge the resultant regions to mitigate the complication caused by dependence. Considering a smaller $n$ is legitimate because when we assume that all 155 cases came from one origin, then any random uniformly produced subset of the 155 cases would come from the same origin, too.  In fact, {\bf W} also considered various subsets of the 155 cases, e.g.\ a subset consisting of 35 cases epidemiologically linked to the Market and a subset consisting of 11 lineage B cases.

The clouds of centroids can be understood as some illustration of the distribution of the estimator of this ``center'', and so can the clouds of modes. However, because the center-points are not rotationally invariant, the clouds of center-points could only be used to provide an idea of the size of variation in the estimator.  Nevertheless, we still show the regions constructed in the same way as the confidence regions.

Visual inspection of Fig.~\ref{fig:4} clearly suggests that the Market (represented by ``{\color{red}${\mathrlap{\raisebox{1.2pt}{$\bigtriangledown$}}{\bigtriangleup}}$}'')
can hardly be considered a part of the clouds of centroids. However, it may be marginally a part of the clouds of modes.  For the clouds of center-points, if we ignore its lack of rotational invariance, when $n \le 100$, the Market could be a part of the clouds of center-points.   In contrast, the Wanda Plaza (represented by ``{\color{magenta}{$\blacksquare$}}'') clearly locates in the central part of the point clouds formed by centroids, and is within the confidence regions of all three kinds of ``center'' when $n=80$.  If the origin of the pandemic is really close to the ``center'' of the point cloud of cases, then in the context of statistics, the Wanda Plaza may be more suspicious than the Market, which is neither more nor less likely to be the origin than the other landmarks shown in the figures are.

To give an argument to justify the Wanda Plaza as an epidemiologically reasonable candidate of the ``center'', we follow {\bf W}, who used the social media check-ins data as geo-tagged information in the years 2013--2014 reported in \citet{LiYangZhuDai2015} to analyse the different patterns of mobility of people for various travel purposes.  Since the link between the number of check-ins and the number of actual visits may vary among different venues \citep[p.~9]{HouLiuNebhenUddinUllahKhan2021}, we have some reservations in the validity of using social media data for the identification of the epicenter.  However, as we mentioned in Section~\ref{sec:2}, the aim of this paper is to criticise not their source of data but only their methods.  Therefore, we also use this data set, in which the check-in places are written in Chinese. However, not all places whose names contain the Chinese word for ``Wanda'' are within the Wanda Plaza area.  A careful inspection of these names and the map of Wuhan City suggested that a lower bound of the count of check-ins within the Wanda Plaza could be obtained by counting places whose names contain the Chinese words for both ``Wanda'' and ``Lingjiao Lake'' (where the Wanda Plaza is next to), and in total 16,317 (out of 770,522) check-ins meet this criterion.  On the other hand, the Market was tagged in 120 check-ins. According to {\bf W}'s Figure~3, a total of 16,317 implies that the Wanda Plaza area, including the Lingjiao Lake,  would be one of the most frequently visited locations throughout Wuhan. If {\bf W}'s use of these check-ins data was considered valid, then the distribution of these data would point to the Wanda Plaza very strongly. Therefore it could also be sensibly hypothesised as a possible origin of the pandemic, and this hypothesis is not rejected.

\section{Dubious tests} \label{sec:4}
Now we are ready to present our main critique of the crucial weakness of {\bf W}, namely, their inappropriate formulation of the hypotheses and the dubious tests.

For the most important case, which we discuss here in detail,
they write on page~2 of~9 of the main paper: {\it We also investigated whether the December
COVID-19 cases were closer to the market than expected based on an empirical null
distribution of Wuhan's population density [data from WorldPop.org ...], with a
median distance to the Huanan market of 16.11$\,$km.} This is all there is, and it is not said what the
``empirical null distribution'' is. More details are presented only on page~5 in
{\bf W}'s Supplementary Materials, which states:
(lines 14--15) {\it To test whether the December cases were closer to the Huanan market than expected, null distributions were
generated from the population density data [$\ldots$].  {\rm (Line $-6$ to line 1 on next page)}  For each point in each pseudoreplicate the Haversine distance to Huanan was calculated, and the median [$\ldots$] distance to Huanan was calculated for each pseudoreplicate.  The median [$\ldots$] distance between all the early December cases ($n=155$) [$\ldots$] [was] compared to these null distributions.\/}

Speaking in words used in mathematical statistics, {\bf W} used a Monte Carlo test \citep[pp.140-143]{DavisonHinkley1997}.
They generated $r$ artificial patterns of 155 cases by simulating an inhomogeneous Poisson process with intensity function proportional to
population density (weighted by age groups) in Wuhan City. For each pattern they determined the 155 distances from the Market to the simulated individual cases and worked out the median of these numbers. The medians are
denoted by $m_i$ with $i=1$, 2, $\ldots$, $r$. In the peer-reviewed version of {\bf W}, the value of $r$ is missing, but in its preprint version, it was stated that $r=1000$.  There is no further explanation of how the reported $p$-value was obtained, and so naturally we assume that the standard Monte Carlo test was adopted.  That is,
the $r$ simulated medians together with the observed median distance $m_0$
for the 155 cases were ordered in decreasing order, and the Monte Carlo $p$-value would be the rank of $m_0$ in this series divided by $r+1$.

The above procedure does not involve the center-point of the point cloud of the cases but uses the median of 155 distances from cases to the Market as the test statistic.
{\bf W} also considers another test using the distance between the center-point of the 155 cases and the Market as the test statistic, and 1 million points were simulated independently according to the age-weighted population density data to represent the center-points of their null model, in which the number 155 became irrelevant.

Against {\bf W}'s use of this procedure we have three main arguments; the first two are technical and the  third one is fundamental.

First, the simulated patterns are hardly typical patterns of residential addresses of early
infected people in an epidemic, since they tend to be patterns that cover rather large
regions, scattered over the full area of Wuhan City, and hence they are comparatively
more sparse and extended than the observed pattern around the Market and any realistic pattern of case locations of an epidemic.
The model used is perhaps appropriate for a rare disease which is not believed to be contagious.
In particular, it may be useful in the case of an elevated incidence due to some additional
cause such a carcinogenic pollution. The work of \cite{Diggle1990} and  \citet{DiggleRowlingson1994} is a nice example; we will revisit this example in Section~\ref{sec:5}.  However, note that in such a context the source of the pollution is usually known and is therefore not estimated from data.

Second, the hypotheses in {\bf W} were not appropriately formulated.  There are two problems.  One is that in order to reach their conclusion that there is strong evidence that the Market was the epicenter of the early December cases, the null hypothesis would have to be that the Market was not the early epicenter, while the alternative hypothesis should be that the Market was the early epicenter. This construction, if adopted, would be weird and result in a very composite null hypothesis, stating that the early epicenter was a point in Wuhan City other than the Market.  The other problem is that the Market was used in the test statistic $m_0$ defined above but in fact did not play a role in {\bf W}'s null model for simulation.  The test statistic $m_0$ could be interpreted as the ``distance'' between the Market and the point cloud of cases, while the corresponding value $m_i$ in a simulated pattern is the ``distance'' between the Market and data generated proportionally to the population density. Therefore, a small $p$-value obtained from the Monte Carlo test could not suggest the role of the Market in the epidemic but would suggest that the point cloud of cases could not be a point cloud drawn from a distribution proportional to the population density.  Since {\bf W} considered the Market the only candidate for the epicenter, a more statistically appropriate statement of the null hypothesis would be that the Market was the epicenter of the early cases, and the alternative hypothesis should be that the epicenter was somewhere else.  The interested reader is referred to \citet{Strong1980} for an inspiring discussion of the formulation of null hypotheses in different branches of science. It shows that the appropriate choice of null hypotheses is crucial. The paper gives examples of fallacies caused by incorrect choice of null hypotheses. Therefore great care is required.

Still harder is the third point: The result of their test can be predicted without any computer work.  For a contagious disease, including COVID-19 and any viral epidemic, the spatial pattern of case locations should be clustered. If one places clusters of 155 points (clusters following the rule of {\bf W} or even clusters following any realistic stochastic model for an epidemic) randomly in the whole city region of Wuhan, then the probability is very small that a cluster-center falls just to a position close to the Market. This means that the null hypothesis of {\bf W} will almost always be rejected by their test statistic. By the same argument, if one replaces the role of the Market in the test by some other landmark in Wuhan City and repeats
what is done in {\bf W}, the same will happen, namely, rejection of the null hypothesis of {\bf W}. In fact, their model assumption that case locations follow the population density represents the assumption that the disease is not contagious, and almost any test with reasonable power would result in rejection when the observed spatial pattern of case locations is clustered.

Thus, the testing procedure in {\bf W} must be considered unacceptable. It does not support the zoonosis hypothesis.

To enable the Market to play a role in the hypotheses, we consider the null hypothesis that the Market is the ``center'' of the 155 cases, then as we discussed in the previous section, it will also be the ``center'' of any resamples. Here, we present the testing procedure by using the centroid to estimate the ``center'' of a point cloud.  The null hypothesis is rejected if the distance $d_0$ from the observed centroid  (represented by ``{\color{magenta}+}'' in Fig.~\ref{fig:4}) of the 155 cases to the Market (represented by ``{\color{red}${\mathrlap{\raisebox{1.2pt}{$\bigtriangledown$}}{\bigtriangleup}}$}'') is significantly longer than the distances $d_i$, where $i=1$, 2, $\ldots$, $m$, from ``{\color{magenta}+}'' to centroids of $m$ replicates of bootstrap samples of size $n$ from these 155 cases.  The same testing procedure can be applied to the center-points and modes.  The Monte Carlo $p$-values, defined to the rank of $d_0$ in the decreasing series formed by $\{d_0, d_1, \ldots, d_m\}$ divided by $m+1$, where $m=999$, corresponding to the simulation shown in Fig.~\ref{fig:4} are given in Table~\ref{table:1}.  Note that for the clarity of visual inspection, only the 95\% confidence regions are shown in Fig.~\ref{fig:4}, and not the 999 individual centroids, center-points, or modes .  The interested reader may use the R code in the supplementary materials to generate figures with these individual points.

These $p$-values reveal that for $n \ge 150$, if we consider the centroids or {\bf W}'s center-point, the null hypothesis has to be rejected at the 0.05 level.  Even though the Monte Carlo $p$-values for smaller $n$ are larger than 0.05 (and so are the $p$-values for modes), the elliptical shape of the confidence regions shown in Fig.~\ref{fig:4} suggests that using the Euclidean distance as the test statistic, implicitly assuming a circular shape, may lower the power of the test, and the failure of rejecting the null hypothesis could be a type II error.  The small $p$-values for large $n$, together with a visual inspection of Fig.~\ref{fig:4}, indicate that the Market cannot be accepted as the ``center'' of the 155 cases.

\begin{table}
\caption{\label{table:1} Monte Carlo $p$-value for testing whether the Market is the ``center''}
\begin{tabular}{||l|cccc|||}
\hline \hline \hline
``center'' & $n=155$ & $n=150$ & $n=100$ & $n=80$  \\ \hline \hline
centroid & 0.030 & 0.024 & 0.060 & 0.122 \\ \hline
center-point & 0.008 &  0.009 & 0.043 & 0.068  \\ \hline
mode & 0.070 & 0.072 & 0.155 & 0.200 \\
\hline \hline \hline
\end{tabular}
\end{table}

\section{Discussion and conclusion} \label{sec:5}
How should one solve the problem of detecting the origin of an epidemic by means of spatial statistics? When an epidemic of interest is not contagious and has some spatially static sources, modern spatial statistical tools can help pinpoint these sources.

For example,  the papers \citet{Diggle1990} and \citet{DiggleRowlingson1994}, mentioned in Section~\ref{sec:4}, studied a non-contagious epidemic of larynx cancer and used spatial point process modelling to link the home addresses of new cancer patients to the incinerator.  However, this was a circumstance very different from the current study: larynx cancer is not a contagious disease, while COVID-19 is. Moreover, to link the cases to the incinerator, those two papers did not try to show that the incinerator was the ``center'' of the larynx cancer cases but showed that parameters describing the change in the intensity function of larynx cancer cases relative to the incinerator were significant; in fact, it is visually clear in Figure~1 of \citet{Diggle1990}
that the incinerator was not the geometrical ``center'' of the larynx cancer cases. Furthermore, they did not use the population data but the locations of lung cancer cases recorded in the same area to form the null spatial distribution.

Another inspiring example is \citet{Snow1855}, in which a cholera outbreak in London in 1854 was traced to a specific water pump on Broad Street.
He used the Voronoi diagram generated by the locations of water pumps in London and identified that the Voronoi cell containing the highest incidence of cholera was the cell generated by the water pump on Broad Street; the removal of the handle of this water pump was considered the beginning of the end of the cholera outbreak in London.

In these situations, when an epidemic has a localised, non-moving source, statistical methods may identify the source by detecting spatial clusters of cases and ruling out other possible origins. Point process models can be used to evaluate whether the clustering of cases around a suspected source is statistically significant.  Voronoi diagrams can be constructed to determine which source, among all possible sources, is closest to the largest number of cases, suggesting it as the most plausible origin. By combining disease mapping, cluster detection, and spatial statistics, researchers can solve the problem of pinpointing the source of an epidemic.

However, determining the origin is much more difficult if obvious points of origin like water pumps or incinerators do not exist, or if the disease is contagious. More models and tools remain to be developed in the field of spatial epidemiology.  While this paper aims to provide a timely critique of the faults in {\bf W}'s methodology of spatial statistics, we would also like to discuss what possible work is still needed for the analysis of the Wuhan COVID-19 data.

Contagious epidemics and those without a clear point source pose greater challenges.  When there are no obvious physical locations to directly link to cases, more sophisticated statistical methods are required.  The existing epidemiological models typically consider the total numbers of infected people in the region of interest, using differential equations, the SIR and SEIR model, see \citet{YadavAkhter2021} for a review.  The spread of COVID-19, however, demonstrates the need for further progress in modelling an epidemic that evolves from an origin or from multiple origins and propagates across space in complex ways.  Statisticians can then adapt such a model to the topography around the suspected area, the distribution of the population, the public transportation network, among other relevant factors.  For example, \citet{MalikGongMoussawiKornissSzymanski2022} modelled the spread of COVID-19 based on SIR model with mobility trends represented by subway transport data.  Such adapted models, if probabilistic in nature, would enable one to generate simulated patterns of infections in time and space around any suspected origin(s).  When concrete and precise data of the actual infection pattern are available, data-driven approaches like maximum likelihood estimation can be employed to identify cluster centers or the origin(s).

However, such models and the corresponding COVID-19 data do not yet exist and therefore spatial statistics is still unable to solve the problem of detecting the origin of COVID-19 in the way sketched.

We come to a clear conclusion: The analysis in the paper {\bf W} does not give an acceptable argument for the centrality of the Market in the 155 December cases.  Our analysis suggests that the Market as well as some other points in its neighbourhood or some other landmarks like the Wanda Plaza are possible spatial ``centers'' of the cases. Neither {\bf W}'s nor our statistical analysis could be used to support or reject the zoonosis hypothesis.

\section{Acknowledgement}

We thank Adrian Baddeley and the anonymous referees for valuable comments on earlier versions of this paper.

\section{Funding}
The second author is supported by the Research Matching Grant Scheme (RMGS-2022-11-08) from the Research Grants Council of Hong Kong.

\section{Data and Code Availability}
All data and the R code used in this paper are available online as supplementary materials at {\it Journal of the Royal Statistical Society\/}.

\section{Conflict of Interest}
The authors declare no conflict of interest.



\end{document}